\documentclass[aps,prb,twocolumn,superscriptaddress]{revtex4-2}

\usepackage{natbib}
\usepackage{setspace}
\usepackage{bm}
\usepackage{amsmath}
\usepackage{graphicx}

\begin{document}


\title{Intrinsic subthermionic capabilities and high performance of easy-to-fabricate monolayer metal dihalide MOSFETs}



\author{D. Logoteta}
\email{demetrio.logoteta@for.unipi.it}
\affiliation{Dipartimento di Ingegneria dell'Informazione, Universit\`a di Pisa, Via G. Caruso 16, 56126 Pisa, Italy}

\author{J. Cao}
\affiliation{School of Electronic and Optical Engineering, Nanjing University of Science and Technology, Nanjing 210094, China}

\author{M. Pala}
\affiliation{Universit\'e Paris-Saclay, Centre National de la Recherche Scientifique, Centre de Nanosciences et de Nanotechnologies, 91120 Palaiseau, France}

\author{P. Marconcini}
\affiliation{Dipartimento di Ingegneria dell'Informazione, Universit\`a di Pisa, Via G. Caruso 16, 56126 Pisa, Italy}

\author{G. Iannaccone}
\affiliation{Dipartimento di Ingegneria dell'Informazione, Universit\`a di Pisa, Via G. Caruso 16, 56126 Pisa, Italy}



\begin{abstract}
We investigate the design of steep-slope metal-oxide-semiconductor field-effect transistors (MOSFETs) exploiting monolayers of transition metal dihalides as channel materials. With respect to other previously proposed steep-slope transistors, these devices require simplified manufacturing processes, as no confinement of the 2D material is needed, nor any tunneling heterojunction or ferroelectric gate insulators, and only $n$- or $p$-type contacts are demanded. We demonstrate their operation by studying an implementation based on monolayer CrI$_2$ through quantum transport simulations. We show that the subthermionic capabilities of the device originate from a cold-source effect, intrinsically driven by the shape of the band structure of the 2D material and robust against the effects of thermalization induced by the electron-phonon interactions. Due to the absence of a tunneling barrier when the device is switched on, current levels can be achieved that are typically out of reach for tunnel FETs. The device also exhibits excellent scaling properties, maintaining a subthermionic subthreshold swing (SS) up to channel lengths as short as 5 nm.


\end{abstract}


\maketitle


\section{Introduction}
The development of electronic circuitries working with lower energy budgets than currently possible with CMOS devices represents the key step towards the development of complex portable applications, the enabling of a denser integration of components and the design of autonomous systems fully powered by green energy harvesting~\cite{IRDS}.    

The quest for a new generation of energy-efficient devices has followed different paths, leading to the emergence of several proposals and research topics. Over the course of the time, transistors based on band-to-band tunneling~\cite{Seabaugh_IEEEproceedings_2010}, impact ionization~\cite{gopalakrishnan2005}, negative capacitance~\cite{salahuddin2008use}, phase transitions~\cite{nakano2012} and electromechanical switching~\cite{Lee2013} have been introduced, though a clear superiority of a single technology over all the others has still to be demonstrated. 
The common, main goal of these research lines is the reduction of the SS of transistors below the thermionic limit of approximately 60~mV/dec, which would allow to operate digital circuits at lower supply voltages and substantially reduce their power consumption~\cite{Seabaugh_IEEEproceedings_2010}. 

The recent investigations of ``cold-source'' transistors originate, in many ways, from the revolution triggered by the discovery of 2D materials and the opportunity it opened up to leverage tailored band structures. These devices achieve subthermionic SSs by virtue of a band-structure-induced energy filtering, that empties the Boltzmann tail of the Fermi-Dirac distribution of carriers~\cite{Liu2018}. This effect can be engineered by both selecting suitable 2D materials and by tuning the electronic properties of a given monolayer by means of mechanical or chemical methods. A distinctive advantage of cold-source devices with respect to negative-capacitance FETs are their immunity to hysteresis issues, while, compared to tunnel-FETs, they can reach higher levels of ON current and exhibits higher immunity to nonidealities such as traps and band tails~\cite{Logoteta2020}. 


The cold-source transistors proposed to date leverage a lateral confinement~\cite{Logoteta2019} or either vertical~\cite{Qiu2018,Logoteta2020,Tang2021} or lateral~\cite{Liu_2020,Marin2020} heterojunctions to obtain the required energy-filtering effect. However, the fabrication of this kind of structures, even at a proof-of-concept level, still presents significant challenges.
\newline
In this paper, we show that cold-source devices can be designed by adopting a simple MOSFET architecture, with the channel realized in a single 2D material and without the need to enforce any lateral confinement. The fabrication of prototypes with such an architecture is nowadays within the technological know-how of several research laboratories throughout the world~\cite{Shim2017}. 
The operation of the proposed devices relies on the presence in the channel material of narrow isolated bands close to the Fermi level. The electronic transport through these bands results in a filtering of the high-energy components of the current spectrum and directly entails subthermionic SSs. We indentified suitable channel materials in the group of transition metal dihalide monolayers, several of which exhibit few low-energy isolated conduction (e.g. CrI$_2$, CrBr$_2$, FeCl$_2$, NiBr$_2$, NiCl$_2$) or valence bands (e.g. LaBr$_2$, VI$_2$, VBr$_2$, VOCl$_2$, VOBr$_2$), gathering within an energy window of roughly 0.5~eV. Despite a significant electron-polar optical phonon (POP) interaction is expected in these materials~\cite{Lin2014}, which can favor the electron thermalization and impair the cold-source filtering~\cite{Logoteta2020}, the electron-POP coupling can be sufficiently screened by engineering the dielectric environment through the use of high-k oxides~\cite{Sohier2016}.  
\newline
We demonstrate the concept by numerically studying the operation of a $n$-type MOSFET based on a monolayer of CrI$_2$, which, among other possible candidates, exhibits a particularly simple band structure. By combining density functional theory (DFT) and quantum transport simulations, we show that, besides a reduced fabrication complexity, this device exhibits excellent performance, achieving minimum subthreshold swings of $\approx 30$~mV/dec, reaching ON current in excess of 300~$\mu$A/$\mu$m and maintaining a subthermionic behavior for ultra-scaled gate lengths of 5~nm.   

\section{Model}

\begin{figure}
    \centering
	\includegraphics[width=0.85\columnwidth]{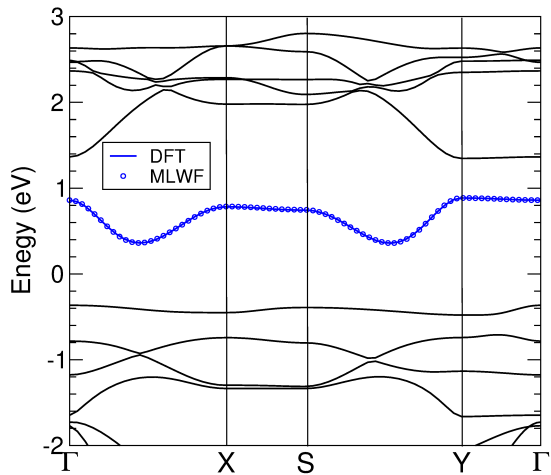}
	\caption{\protect \footnotesize DFT band structure of monolayer CrI$_2$. The isolated band considered in transport simulations is highlighted in blue. The circles refer to the same band within the MLWF tight-binding model. The Fermi level is set at 0 eV.}  
  \label{fgr:bands}
\end{figure}

Monolayer CrI$_2$ is a polymorph of the already intensively studied CrI$_3$~\cite{Huang2017}. It consists of a plane of chromium atoms stacked between two planes of iodine atoms and is obtainable by direct exfoliation of its parent 3D material~\cite{mounet2018two}. In this paper, we refer to its 1-T antiferromagnetic phase, predicted to be the most energetically stable~\cite{Kulish2017}. 

\begin{figure}
    \centering
	\includegraphics[width=1.0\columnwidth]{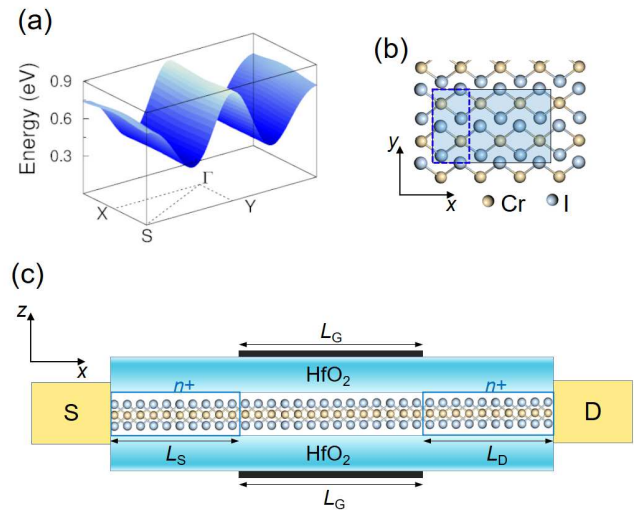}
	\caption{\protect \footnotesize (a) Plot of the lowest conduction band of monolayer CrI$_2$ over the first Brillouin zone. (b) Primitive unit cell of monolayer CrI$_2$ (dashed rectangle) and the supercell considered in transport simulations (solid rectangle). (c) Sketch of the transistor studied in the paper.}
  \label{fgr:Sketch}
\end{figure}

According to the DFT band structure shown in Fig.~\ref{fgr:bands}, monolayer CrI$_2$ is a semiconductor, with a band gap of approximately 0.7~eV. The lowest conduction band, highlighted in blue, exhibits a small width of around 0.52~eV and a marked anisotropy: the effective mass is equal to $m_x=0.5$~$m_0$ along the $\Gamma$-X direction and $m_y=5.6$~$m_0$ along the orthogonal $\Gamma$-Y direction (see Fig.~\ref{fgr:Sketch} (a)).  
We restrict ourselves to consider only this band in our analysis, since the large separation of about 0.5 eV from the next higher-energy band  makes any other contribution to the electron conduction negligible. To simulate the transport, the DFT Hamiltonian was mapped into a single-orbital tight-binding model, obtained by projection over a basis of maximally localized Wannier functions (MLWF). The excellent agreement with the DFT results is shown in Fig.~\ref{fgr:bands}. Moreover, in order to take into account all the significant couplings beyond the nearest-neighbor approximation, the CrI$_2$ unit cell was expanded into a supercell including three primitive cells (see Fig.~\ref{fgr:Sketch} (b)).


Our transport model incorporates the electron interactions with acoustic and polar optical phonons. The latter are expected to provide the dominant inelastic scattering mechanism, due to the significant ionic character of transition metal halides~\cite{Lin2014}. We describe the electron-phonon coupling within a DFT-based deformation potential approximation, which also takes into account the screening of phonons by the dielectric environment. 

In order to obtain a fully self-consistent solution, the transport equations have been nonlinearly coupled with the Poisson equation. 

More specific details about the model and the simulation strategy are provided in the Method section and in the Supplementary Material.

\section{Results and Discussion}

\begin{figure*}
    \centering
	\includegraphics[width=1.4\columnwidth]{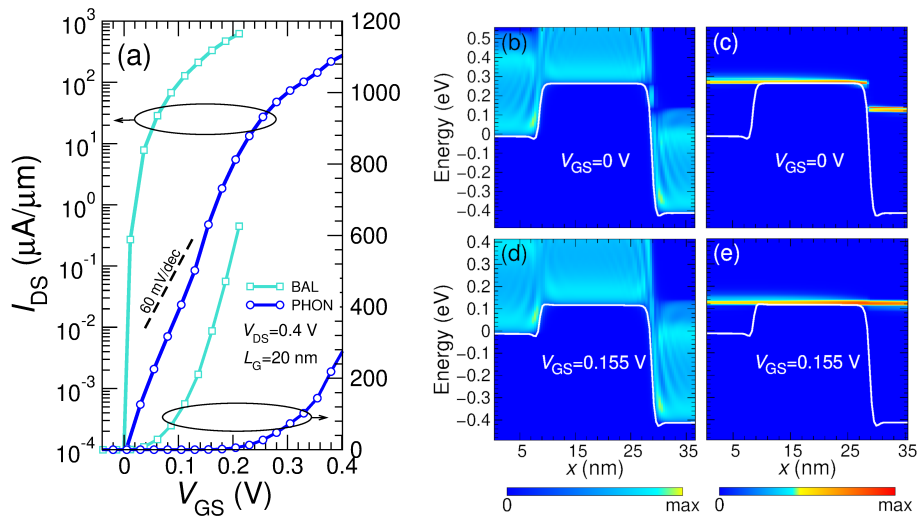}
	\caption{\protect \footnotesize (a) Transfer characteristics of the CrI$_2$-based MOSFET for $L_{\rm G}=20$~nm and $V_{\rm DS}=0.4$~V, in the ballistic transport regime (BAL) and in the presence of electron interactions with acoustic and polar optical phonons (PHON). The plot of the ballistic transfer characteristics is restricted to the window in which a reliable convergence of the simulation was obtained. (b) Local  density of states and (c) current spectrum along the device at $V_{\rm GS}=0$~V. (d) Local  density of states and (e) current spectrum along the device at $V_{\rm GS}=0.155$~V. The bottom edge of the conduction band is indicated by the white curves. The Fermi level at the source contact is set at 0~eV.}  
  \label{fgr:IV_ballistic}
\end{figure*}

The transistor studied in the paper is sketched in Fig.~\ref{fgr:Sketch} (c). The transport direction is assumed to be along the $x$ axis. We consider a double-gate architecture with gates of the same length as the channel and 3~nm-thick HfO$_2$ insulator layers. The source and drain extensions of the CrI$_2$ channel are chemically doped at a concentration of $5\times 10^{13}$ cm$^{-2}$ and have a length of $L_{\rm S}=L_{\rm S}\simeq$ 8~nm. The channel is undoped.  

Fig.~\ref{fgr:IV_ballistic} (a) shows the transfer characteristics of the transistor for a gate length $L_{\rm G}=20$~nm and a source-to-drain bias $V_{\rm DS}=0.4$~V. The ballistic characteristics, obtained by setting to zero the electron-phonon coupling, are also reported. By following the IRDS projections for low-power applications~\cite{IRDS}, here and throughout the paper, we define as the OFF-state voltage the value of $V_{\rm GS}$ at which the current $I_{\rm DS}$ equals the value $I_{\rm OFF}=10^{-4}$ $\mu$A/$\mu$m. For the sake of clarity, all the transfer characteristics are shifted in order to set $V_{\rm GS}^{\rm OFF}=0$ V. The ON-state voltage is computed as $V_{\rm GS}^{\rm OFF}+V_{\rm DS}$. 

The device exhibits an average SS of about 43~mV/dec over the window $I_{\rm OFF} < I_{\rm DS} < 1$ $\mu$A/$\mu$m, and a minimum SS of $\approx 33$~mV/dec. The ON current amounts to $I_{\rm ON}=285$~$\mu$A/$\mu$m, a value significantly larger than those typically obtained with similar biases in tunnel FETs~\cite{Lu_JEDS_2014}. This improvement originates from the absence of any tunneling barrier in the ON state, which is the main cause of the current degradation in tunnel-FETs. The limiting factor of $I_{\rm ON}$ in the considered device is instead the significant back-scattering from acoustic phonons, sustained by the high density of states ($\propto \sqrt{m_x m_y}$ close to the band edge) in CrI$_2$. Its effect on the device performance can be quantitatively appreciated in Fig.~\ref{fgr:IV_ballistic} in terms of the large decrease of the above-threshold current with respect to the regime of ballistic transport.

 
To explain the occurrence of subthermionic SSs, we plot in Fig.~\ref{fgr:IV_ballistic} (b) to (e) the local density of states (LDOS) and the current spectrum along the device at different gate voltages. Due to the narrowness of the conduction band, at low enough $V_{\rm GS}$ the electron states in the channel do not overlap in energy with the states in the drain (Fig.~\ref{fgr:IV_ballistic} (b)). Electrons injected from the source contact are thus mostly reflected back. The residual leakage current is due to a phonon-assisted relaxation (thermalization) of the electrons in the channel toward the states at lower energy in the drain (Fig.~\ref{fgr:IV_ballistic} (c)) and, to a lesser extent, to the source-to-drain tunneling through the channel barrier. Consistently, the ballistic transfer characteristics in Fig.~\ref{fgr:IV_ballistic} (a), in which the effect of relaxation is removed, exhibit much smaller SS values of less than 5~mV/dec. 
For the considered gate length of~20 nm and the considered $V_{\rm GS}$ window, electron relaxation is the dominant leakage mechanism. Source-to-drain tunneling can play a more significant role at shorter gate lengths, as will be discussed later.   

As $V_{\rm GS}$ increases, the energy gap between the states in the channel and in the drain becomes gradually smaller, until an overlap is established at $V_{\rm GS}\approx 0.155$ V (Fig.~\ref{fgr:IV_ballistic} (d) and (e)). This is also the approximate $V_{\rm GS}$ value at which the SS attains its minimum. 
  
\begin{figure}
    \centering
	\includegraphics[width=0.7\columnwidth]{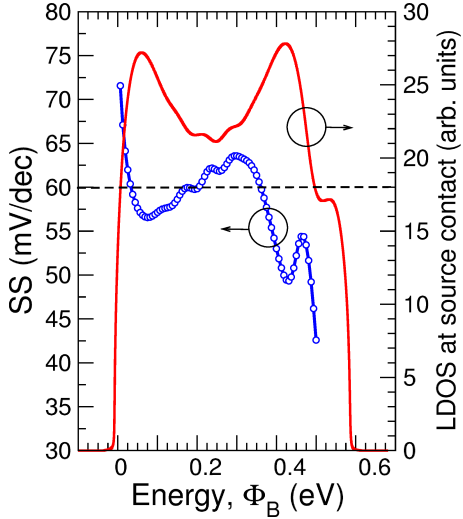}
	\caption{\protect \footnotesize Local density of states (LDOS) at the source contact and SS as a function of the height of the barrier $\Phi_B$ computed from Eq.~(\ref{eq:SS}). $V_{\rm DS}=0.4$~V, $L_{\rm G}=20$~nm.}  
  \label{fgr:SS}
\end{figure}

Besides the modulation of the electron transmission from the channel to the drain contact, another minor effect contributes to the subthermionic capabilities of the device. Its origin can be identified more clearly by assuming ballistic transport and a step-like dependence of transmission on energy. Specifically, we assume zero transmission at energies below the top of the barrier and generically constant transmission above it. Under these hypotheses, the transport can be described by the Landauer formula~\cite{Meir1992} and the subthreshold swing can be expressed in the form (see the Supplementary Information for a derivation): 

\begin{multline}
SS = \frac{\partial V_{\rm G}}{\partial \log_{10} I_{\rm DS}} = \frac{KT}{q}\ln(10)\left(\frac{\partial \Phi_B}{\partial (q\,V_{\rm GS})} \right)^{-1}  \\ \times\left( 1+\frac{\int_{\Phi_B}^\infty dE e^{-E/KT} \frac{\partial D(E)}{\partial E}}{e^{-\Phi_B/KT}D(\Phi_B)}\right) \tag{1}\label{eq:SS}, 
\end{multline}
where $K$ is the Boltzmann constant, $T$ is the room temperature, $q$ is the absolute value of the electron charge, $\Phi_B$ is the height of the channel barrier with respect to the source Fermi level, $D(E)$ is the local density of states at the source contact, and $E$ is the energy. Eq.~(\ref{eq:SS}) represents a generalization of a similar expression found, in the specific case of graphene, in Ref.~\cite{Qiu2018}.

The product $\frac{KT}{q}\ln(10)\left(\frac{\partial \Phi_B}{\partial (q\,V_{\rm GS})} \right)^{-1}$ equals 60~mV/dec if the gate control over the barrier is perfect, namely if $\left(\frac{\partial \Phi_B}{\partial (q\,V_{\rm GS})} \right)^{-1}=1$. The SS can attain subthermionic values if the remaining factor on the rightmost side of Eq.~(\ref{eq:SS}) is smaller than 1, which translates to the condition $\int_{\Phi_B}^\infty dE e^{-E/KT} (\partial D(E)/\partial E) < 0$. This constraint requires $\partial D(E)/\partial E$ to be negative at least over some finite energy intervals above the top of the barrier. To show that this is indeed the case for the device at hand, we plot in Fig.~\ref{fgr:SS} the actual LDOS (including the broadening due to the electron-phonon scattering) at the source contact and the SS value computed from Eq.~(\ref{eq:SS}). The SS is subthermionic over two subintervals, approximately corresponding to the windows where the derivative of the LDOS with respect to the energy is negative. Within the range of $V_{\rm GS}$ of interest here, the device can benefit from this effect only in the subinterval 0.01--0.2 eV, where the condition $\left(\frac{\partial \Phi_B}{\partial (q\,V_{\rm GS})} \right)^{-1}=1$ is still verified with good approximation and the SS attains a minimum value of 56~mV/dec. In the case at hand, the improvement of the SS attributable to this effect is therefore quite modest. Nevertheless, as suggested in Ref.~\cite{Qiu2018}, the principle is general and optimized materials where it can be emphasized can be found. 

\begin{figure}
    \centering
	\includegraphics[width=0.8\columnwidth]{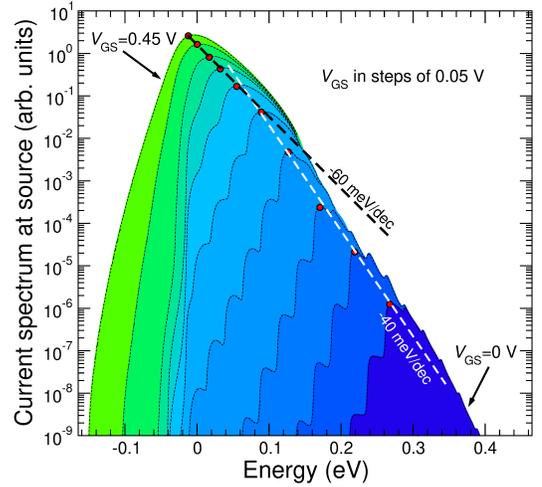}
	\caption{\protect \footnotesize Current spectrum at the source contact for ten evenly-spaced $V_{GS}$ values in the range [0-0.45]~V. The red circles indicate the maximum of the spectra. $V_{\rm DS}=0.4$~V, $L_{\rm G}=20$~nm.}   
  \label{fgr:Spectrum}
\end{figure}

The occurrence of a cold-source effect, resulting from the combination of the previously discussed physical mechanisms, can be visualized in Fig.~\ref{fgr:Spectrum} in terms of the evolution of the current spectrum at the source contact. 
At high $V_{\rm GS}$ values, the maxima of the current spectra, roughly coinciding with the centroid of the same, are interpolated by a -60~mV/dec straight line, as expected in a normal MOSFET. However, at low enough $V_{\rm GS}$, the maxima approximately lie over a steeper -40~mV/dec line, in agreement with the SS value computed from the transfer characteristics.     



\begin{figure}
    \centering
	\includegraphics[width=1.0\columnwidth]{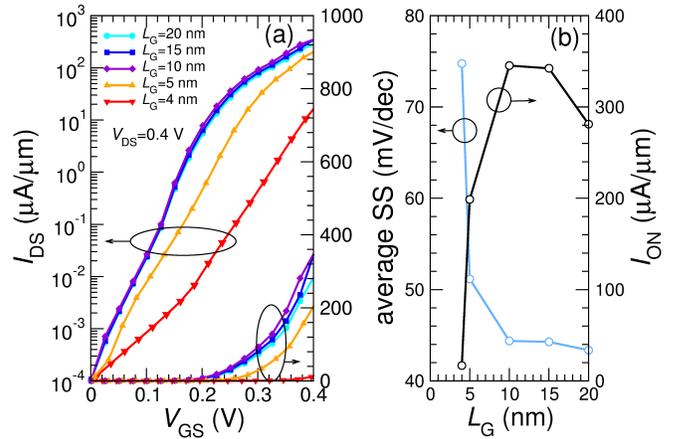}
\caption{\protect \footnotesize (a) Transfer characteristics of the transistor for different values of $L_G$. (b) Average SS over the window $10^{-4}$ $\mu$A/$\mu$m $< I_{\rm DS} < 1$ $\mu$A/$\mu$m and ON current as a function of $L_G$.}
  \label{fgr:LG}
\end{figure}

The scaling behavior of the device is investigated in Fig.~\ref{fgr:LG}. As expected in a MOSFET architecture, the progressive degradation of the SS as $L_{\rm G}$ is decreased is due to the increment of the source-to-drain tunneling. Remarkably, the device exhibits subthermionic capabilities in the $V_{\rm GS}$ range of interest for gate lengths as short as 5~nm. This is to be traced back to the relatively high value of the effective mass $m_x$ in the transport direction, that effectively suppresses the tunneling through the barrier. 

\begin{figure}
    \centering
	\includegraphics[width=1.0\columnwidth]{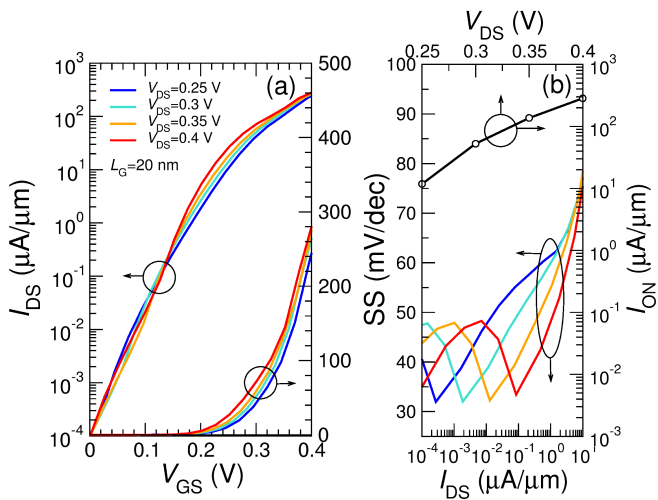}
	\caption{\protect \footnotesize (a) Transfer characteristics of the transistor for different values of $V_{\rm DS}$. (b) SS as a function of the current for the same set of $V_{\rm DS}$ values and ON current as a function of $V_{\rm DS}$.}    
  \label{fgr:VDS}
\end{figure}

The operation of the device at smaller $V_{\rm DS}$ is explored in Fig.~\ref{fgr:VDS}. The main effect on the transfer characteristics of reducing $V_{\rm DS}$ is a gradual shift toward lower current levels of the steepest part of the curves (see Fig.~\ref{fgr:VDS} (b)). 
This behavior can be easily understood by considering that the onset of the overlap between the energy states in the channel and in the drain roughly occurs when the barrier height is $\Phi_B\approx \Delta E_c -qV_{\rm DS}$, where $\Delta E_c$ is the width of the conduction band. Thus, as $V_{\rm DS}$ decreases, the overlap arises at higher $\Phi_B$ and, consequently, at lower $V_{\rm GS}$. 
This effect progressively degrades the average SS of the device over the window of interest for digital low-power applications ($I_{\rm DS}\geq 10^{-4}$~$\mu$A/$\mu$m), thus contributing to deteriorate the ON current when the source-to-drain bias is reduced (Fig.~\ref{fgr:VDS} (b)). 


\begin{figure}
    \centering
	\includegraphics[width=0.8\columnwidth]{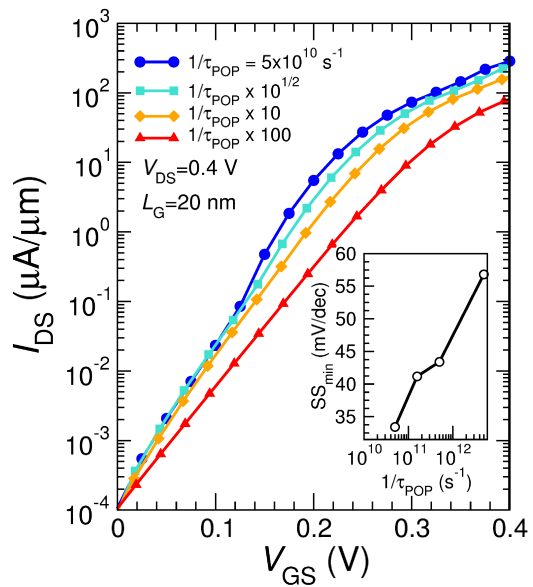}
	\caption{\protect \footnotesize Transfer characteristics of the transistor when the electron-POP scattering rates are rescaled by a factor of 1, $\sqrt{10}$, 10 and 100. The numerical value of $1/\tau_{\rm POP}$ reported in the legend refers to the estimated phonon absorption rate at the conduction band minima. {\it Inset}: minimum of SS as a function of the electron-POP scattering rate.}   
  \label{fgr:POP}
\end{figure}

We conclude by considering in more detail the inelastic electron-phonon interaction. According to the previous discussion, it can importantly impact the performance of the transistor, due to its significant contribution in determining the subthreshold leakage current. 
It is of particular interest exploring the device operation for stronger electron-POP couplings than assumed in the previous analysis, in order to assess the device sensitivity to the value used for the POP deformation potential. The transfer characteristics in Fig.~\ref{fgr:POP} show that, in order to completely suppress the subthermionic effect, the electron-POP scattering rates have to be increased by more than two order of magnitude with respect to the specific value estimated for our system ($1/\tau_{\rm POP}=5\times 10^{10}$~s$^{-1}$ for the phonon absorption rate at the minima of the conduction band; see the Supplementary Material for the details of the calculation). A significant margin thus exists, which suggests that the operation of the device is robust enough against the cumulative effect of minor electron-phonon interactions (e.g. associated to non-polar optical branches) not considered in this paper. We finally notice that the electron-POP scattering rates could in principle be further reduced by embedding the transistor channel in a dielectric environment with higher permittivity than HfO$_2$~\cite{Sohier2016}. By following this approach, an improvement of the device performance can be realized as long as electron-POP interactions remain the dominant source of inelastic scattering.



\section{Method}
DFT simulations of monolayer CrI$_2$ were performed in a plane-wave basis by means of the Quantum ESPRESSO suite~\cite{Giannozzi2009,Giannozzi2017}. The DFT Hamiltonian was obtained through a spin-polarized calculation on a $12\times12\times1$ Monkhorst-Pack $k$-point grid, by using a Perdew-Burke-Ernzerhof exchange-correlation functional and by assuming a cutoff energy of 100 Ry. Van der Waals interactions were taken into account through the DFT-D3 Grimme's corrections~\cite{grimme2010consistent}. The DFT Hamiltonian was mapped into a single-orbital tight-binding form by projection over a basis of maximally localized Wannier functions through the wannier90 package~\cite{Mostofi2014}. A supercell consisting of three primitive cells and thus including couplings up to third neighbors was used for the transport simulation. 

The value of the static and high-frequency dielectric permittivity tensors, and of the Born effective charge tensors were computed in the framework of the Density Functional Pertutbation Theory through the ph.x and EPW~\cite{PONCE2016} codes. The contribution of the vacuum layer to the values computed for the permittivity was removed by following the method described in Ref.~\cite{laturia2018dielectric} and assuming a CrI$_2$ monolayer thickness of 0.657~nm. 
The relatively small anisotropy in the in-plane static and high-frequency dielectric tensors (less than 10\%) was neglected and a single average value of $\epsilon_{\parallel}^{\infty}\simeq 6.5$ and $\epsilon_{\parallel}^0\simeq 8.5$ was used. The same isotropic approximation was applied to the Born effective charge tensors. The out-of-plane value of the dielectric tensor was evaluated to be $\epsilon_{\perp}^{\infty}=\epsilon_{\perp}^0\simeq 5$. 

The electron-acoustic phonon coupling was described within a deformation potential approximation. 
The acoustic deformation potential was computed as $D_{\rm AC}=\partial E_c/\partial\delta\approx 2$~eV, by fitting the slope of the function $E_c(\delta)$, where $E_c$ is the energy of the bottom edge of the conduction band and $\delta$ is an applied biaxial strain along the axes of the unit cell. 

The electron-POP coupling was described through an effective deformation potential, calibrated on the absorption scattering rate at the minima of the conduction band. The value of the scattering rate $1/\tau_{\rm POP}$ was evaluated within the framework of the theory developed in Refs.~\cite{Verdi2015,Sohier2016}, through the following formula (see the Supplementary Information for more details): 

$$\frac{1}{\tau_{\rm POP}}=\frac{n_{\hbar\omega} q^4\hbar}{4A\left(\epsilon_0 r_{\rm eff}\right)^2\left(\hbar\omega\right)^2 N_{\rm at}} \left(\sum_a \frac{|Z_a|}{\sqrt{M_a}}\right)^2\,\, ,$$ 

where $\epsilon_0$ is the vacuum permittivity, $\hbar\omega$ is the energy of the polar optical phonons, $n_{\hbar\omega}$ is the value of the Bose-Einstein distribution at energy $\hbar\omega$, $N_{at}$ is the total number of atoms in the unit cell, $Z_a$ and $M_a$ are the in-plane effective Born charge and the mass of the atom $a$ in the unit cell, respectively, and $A$ is the area of the unit cell. The parameter 
$r_{\rm eff}$ accounts for the screening from both the monolayer and the top and the top and bottom gate insulators. Its definition is provided in Ref.~\cite{Sohier2016}.


The value of the effective electron-POP  deformation potential $D_{\rm POP}=0.4\times 10^8$~eV/cm was derived from the equation 
$$\frac{D_{\rm POP}^2}{\rho\omega}\left(\frac{\sqrt{m_x m_y}}{2\pi\hbar^2}\right)n_{\hbar\omega}= \frac{1}{\tau_{\rm POP}}\,\, ,$$ 
where the left-hand side is the expression of the absorption scattering rate at the conduction band minima within the self-consistent Born approximation and $\rho$ is the mass density of monolayer CrI$_2$.


Transport simulations were performed within the framework of the non-equilibrium Green's function formalism by using a home-made code. The device was modeled as translationally invariant in the direction $y$ orthogonal to transport and a set of 21 transverse wave vectors $k_y$ were used to sample the Brillouin zone. Transport equations were self-consistently solved with the Poisson equation in the device cross-section. Electron-phonon interactions were included within the self-consistent Born approximation, by assuming an elastic and dispersionless approximation for acoustic and polar optical phonons, respectively.

\section{Conclusion}
We have proposed the design of steep-slope MOSFETs based on transition metal dihalides. We have demonstrated the concept by studying a transistor in which a monolayer of CrI$_2$ serves as channel. The subthermionic capabilities of the device are driven by a cold-source effect, intrinsically entailed by the band structure of the 2D material and protected from a thermalization-driven weakening by a suitable engineering of the dielectric environment of the channel. The structure of the device is particularly simple, as it does not require any confinement or the presence of heterojunctions along the channel and only needs $n$- or $p$-type contacts. The absence of tunneling barriers in the ON state allows the achievement of higher values of current than typically possible for tunnel FETs, while the relatively high value of the transport effective mass enables an aggressive scaling. 

Similar cold-source effects could be observed in other semiconducting monolayers of metal dihalides, which share with CrI$_2$ the presence of isolated and narrow conduction or valence bands close to the Fermi level. We thus suggest that this class of monolayers represents a promising corner of the 2D materials world to explore in the search of candidates and opportunities for the development of a post-CMOS, low-power electronics.

\bibliography{biblio}

\end{document}